\documentclass[fleqn,12pt,twoside]{article}
\usepackage{espcrc1}

\usepackage{graphicx}
\usepackage[figuresright]{rotating}

\newcommand{\AmS}{{\protect\the\textfont2
  A\kern-.1667em\lower.5ex\hbox{M}\kern-.125emS}}

\title{Small x physics and the initial conditions in heavy ion collisions}

\author{Alex Krasnitz\address{UCEH, Universidade do Algarve,\\
        Campus de Gambelas, P-8000 Faro, Portugal.}
 and Raju Venugopalan\address{Physics Department and RIKEN-BNL Research 
Center,\\ 
        Brookhaven National Laboratory, Upton, NY 11973, USA}%
        \thanks{R. V.'s research was supported by DOE Contract
No. DE-AC02-98CH10886.  The authors acknowledge support from the
Portuguese FCT, under grants CERN/P/FIS/1203/98 and
CERN/P/FIS/15196/1999.}
        }
       
\begin{document}

\maketitle

\begin{abstract}

At very high energies, the high parton densities (characterized by a
semi-hard saturation scale $\Lambda_s$) ensure that parton
distributions can be described by a classical effective field theory
with remarkable properties analogous to those of spin glass
systems. This color glass condensate (CGC) of gluons also provides the
initial conditions for multi-particle production in high energy
nuclear collisions.  In this talk, we briefly summarize recent
theoretical and phenomenological progress in the CGC approach to small
x physics. In particular, we discuss recent numerical work on the real
time gluodynamics of partons after a nuclear collision. The
implications of this work for the theoretical study of thermalization
in nuclear collisions and on the phenomenological interpretation of
results of the recent RHIC experiments are also discussed.

\end{abstract}

\section{Introduction}

Why is small $x$ physics relevant to heavy ion collisions? The
canonical picture of how matter is produced and evolves in a high
energy nuclear collision was articulated by
Bjorken~\cite{Bjorken}. The ``valence'' components of the nuclear
wavefunction (partons carrying a large fraction, per nucleon, of the
nuclear momentum) interact weakly and populate primarily the
fragmentation regions at large projectile and target rapidities. It is
the ``wee'' virtual excitations of the nuclear ground state, gluons
(mostly) and sea quarks, carrying a small fraction of the nuclear
longitudinal momentum, that populate the central rapidity region,
eventually producing the $\sim 1000$ hadrons that were measured in one
unit of pseudorapidity at RHIC.  If the nucleus is viewed as a
superposition of Fock states, these wee partons belong to a state with
a large number of gluons and sea quarks--each of these can therefore
have only a small fraction of the nuclear momentum~\cite{BjKoSo}.  The
higher the energy of the nucleus, the larger the number of wee
partons. In a nuclear collision, this Fock state is ``destroyed'' and
the wee partons go on--shell creating the large energy densities in
the relatively large space--time volumes associated with quark gluon
plasma formation.  Understanding the possible formation and dynamical
evolution of this partonic state therefore requires understanding
small $x$ physics at a deep level.

In the perturbative~\cite{BFKL} QCD approach to small $x$ physics, the
``cloud'' of gluons is generated by Bremsstrahlung of a laddder of
soft gluons off harder ``valence'' partons. The parton density grows
rapidly with decreasing $x$ at a rate which, if extrapolated to
$x\rightarrow 0$, would violate the Froissart (unitarity) bound. This
seems unlikely because at sufficiently small $x$ the Bremsstrahlung
approach must break down. When the density of partons becomes large,
parton recombination and screening effects become important, and may
perturbatively ``saturate'' the rapid growth of parton distributions
at small $x$~\cite{GLRMQ}.

Since the density of partons is large, one can formulate the small $x$
problem as a classical effective field theory (EFT)~\cite{MV}~
\footnote{The approach described here is by no means the only
approach. For a review of the recent status of different approaches to
small $x$ physics, see Ref.~\cite{Pramana}.}. The small $x$ partons
described by the EFT constitute a {\it color glass condensate}
(CGC)~\cite{RajGavai,ILM}.  They are a glass because their properties
are analogous to a spin glass. They form a condensate because
they belong to a state with high occupation number of order
$1/\alpha_S$.

The CGC contains a dimensionful scale, the saturation scale
$\Lambda_s(x,A)$. It is proportional to the gluon density per unit
area and grows as $\Lambda_s^2\sim A^{1/3}/x^\delta$, where
$\delta\approx 0.2-0.3$ is the power that controls the rise in parton
distributions in a nucleon. If $\Lambda_s \gg \Lambda_{QCD}$, weak
coupling methods are applicable and one can compute the parton
distributions in the classical EFT. At RHIC, one can estimate
$\Lambda_s= 2$ GeV, while at LHC, $\Lambda_s=4$ GeV --weak coupling
methods are therefore only marginally applicable at the
former~\cite{GyMcl}.

This talk is organized as follows. In section 2, we discuss the color
field of a single nucleus. We will advertise the impressive recent
theoretical developments in this approach. We will also discuss the
striking phenomenological success of saturation models in
describing the HERA data.  In the following section, we will discuss
the application of these ideas to nuclear scattering.  Results from
non--perturbative numerical simulations of the initial stages of
nuclear collisions will be presented. In section 4, we will briefly
confront saturation models against the recent RHIC data. We end in
section 5 with an outlook describing relevant projects in progress.

\section{The color field of a nucleus: QCD is a color glass condensate 
at high energies}

The classical EFT approach to small $x$ physics has been discussed
extensively in the literature~\cite{MV,RG,ILM}. It contains a
dimensionful scale $\Lambda_s$ equal to the color charge squared per
unit area of the valence sources. The EFT, formulated in the infinte
momentum frame, in light cone gauge, is a (nearly) two-dimensional 
theory with a structure mathematically analogous to that of a
disordered system of Ising spins in a random magnetic
field~\cite{ParisiSourlas,RajGavai}.  That's where the color glass
comes from~\footnote{Unlike the original model of Parisi and Sourlas,
which fails in describing real spin glasses, this theory, on account
of its peculiar quantum structure, may indeed be a glass. This remains
to be proven though.}. Further, since the occupation number of the
gluonic fields is large--the field strength squared $F_{\mu\nu}^2\sim
1/\alpha_S$--what we have is a color glass condensate (CGC)
~\cite{ILM}. A virtual photon (from an energetic leptonic probe) would
therefore indeed ``see'' a colored glass on the time scales of its
interaction.

The classical theory is solvable and classical correlation functions
have been computed~\cite{LarKov}. The gluon number has the
Weizs\"acker--Williams distribution $\Lambda_s^2/k_t^2/\alpha_S$ for
$k_t>>\Lambda_s$ and (due to an infinite resummation in
$\Lambda_s/k_t$) saturates as the distribution
$\ln(\Lambda_s/k_t)/\alpha_S$. This suggests that the typical momentum
of the gluons is peaked around $\Lambda_s$. Recently, Lam and Mahlon
have done a lot of work elucidating the nature of the classical
distributions when strict color neutrality is imposed~\cite{Lam}.

When  quantum corrections are included, a 
Wilsonian renormalization group (RG) picture emerges. The structure of 
the classical field remains intact while the scale $\Lambda_s$ grows. 
This is because including hard gluons (harder than the $x$ scale of interest) 
in the source increases the typical size of color fluctuations. The scale 
$\Lambda_s(x)$ is then given by the gluon density per unit area, 
\begin{equation}
\Lambda_s^2 (x) = \frac{A^{1/3}}{x^\delta}.
\end{equation}
Here $\delta(\Lambda_s)$ denotes the power of the rise in the {\it nucleon} 
gluon distribution. This equation must be solved self--consistently for 
$\Lambda_s$ at each $x$. 

In particular, the weight function that gluonic correlation functions
have to be averaged over satisfies a non--linear RG
equation~\cite{RG}. There has been very important work recently in
understanding and {\it solving} this RG equation~\cite{ILM}.  (The
correlation functions obtained from this functional equation are
identical to those obtained independently in different approaches by
Balitsky and Kovchegov~\cite{BalKov}. For numerical solutions of these, see
Ref.~\cite{BraunLev}.) The equations have been solved
in a mean field approximation--for large transverse momenta ($k_t>>
\Lambda_s$) the structure of the source is the simple Gaussian one of
the McLerran--Venugopalan model, while at small transverse momenta, ($k_t<<
\Lambda_s$) it is entirely different. Much work remains to be done to 
uncover the complex structure of these equations--their phenomenological 
implications for heavy ion collisions and for deeply inelastic 
scattering~\cite{eRHIC} are tremendous.

Simple saturation models which incorporate the scale $\Lambda_s$ have
been applied to study the HERA data. With only three parameters, they
fit the entire range of the HERA inclusive data (for $x<0.01$ and for
all $Q^2$ upto $450$ GeV$^2$). The model, with the same parameters, 
reproduces the HERA diffractive and (with some caveats) vector meson production
data~\cite{Golec}.

\section{Classical approach to nuclear collisions: shattering the color 
glass condensate}

In a nuclear collision, the color glass shatters producing a large
number of gluons. The full quantum description of how this happens is
a very difficult problem which has no solution yet. The standard pQCD
factorization (``mini-jets'') approach to particle production is not
of much help here because it is most unreliable at small $x$ and
small $Q^2$.  Inevitably, the better pQCD based models of nuclear
scattering include saturation ideas in some form~\cite{EKRT}.

However, one can make quite a bit of headway in the classical EFT
approach.  Since the classical fields of the two nuclei are known,
they provide the initial conditions for the evolution of the gluon
fields produced in the nuclear collision.  The small $x$ fields are
described by the classical Yang-Mills equations
\begin{equation}
D_\mu F_{\mu\nu}=J_\nu\label{eqmo}
\end{equation}
with the random sources on the two light cones:
$J_\nu=\sum_{1,2}\delta_{\nu,\pm}\delta(x_\mp)\rho_{1,2}(r_t).$ The
two signs correspond to two possible directions of motion along the
beam axis $z$. As shown by Kovner, McLerran and Weigert
(KMW)~\cite{KMW}, low-$x$ fields in the central region of the
collision obey sourceless Yang-Mills equations (this region is in the
forward light cone of both nuclei) with the initial conditions in the
$A_\tau=0$ gauge given by
\begin{equation}
A^i=A^i_1+A^i_2;\ \ \ \ A^\pm=\pm{{ig}\over 2}x^\pm[A^i_1,A^i_2].
\label{incond}
\end{equation}
Here the pure gauge fields $A^i_{1,2}$ are solutions of (\ref{eqmo})
for each of the two nuclei in the absence of the other nucleus.

In order to obtain the resulting gluon field configuration at late
proper times, one needs to solve (\ref{eqmo}) with the initial
condition (\ref{incond}).  Since the latter depends on the random
color source, averages over realizations of the source must be
performed.  KMW showed that in perturbation theory the gluon number
distribution by transverse momentum (per unit rapidity) suffers from
an infrared divergence.  A reliable way to go beyond perturbation
theory is to re-formulate the EFT on a lattice by discretizing the
transverse plane. (Boost invariance in pseudorapidity is assumed, which 
reduces the theory to a $2+1$--dimensional theory.)
The resulting lattice theory can then be solved
numerically~\footnote{For alternative analytical approaches, see the 
work of Balitsky~\cite{Ian} and of Kovchegov~\cite{Yura}.}. We shall not dwell here on the details of the lattice
formulation, which is described in detail in Ref.~\cite{AlexRaj1}.

We have simplified the problem considerably since $\Lambda_s$ and the
linear size $L$ of the nucleus are the only parameters in the problem.
We can write any dimensionful physical quantity $q$ as
$\Lambda_s^df_q(\Lambda_s L)$, where $d$ is the dimension of $q$. All
the non-trivial physical information is contained in the dimensionless
function $f_q(\Lambda_s L)$~\cite{RajGavai}.  We can estimate the
values of the product $\Lambda_s L$ which correspond to key collider
experiments. Assuming Au-Au collisions, we take $L=11.6$ fm (for a
square nucleus!) and estimate the standard deviation $\Lambda_s$ to be
2 GeV for RHIC and 4 GeV for LHC~\cite{GyMcl}.  Also, we
have approximately $g=2$ for energies of interest. The rough estimate
is then $\Lambda_s L\approx 120$ for RHIC and $\Lambda_s L\approx 240$
for LHC.

Our simulations were performed for the SU(2) gauge group--all results
for physical quantities are based on extrapolations to SU(3).  A
straightforward computation is of the initial energy per unit transverse area
per unit rapidity, deposited in the central region by the colliding
nuclei~\cite{AlexRaj2}. At late times, this is given by the non--perturbative 
formula 
\begin{equation}
\frac{1}{\pi R^2}\,\frac{dE_t^i}{d\eta}|_{\eta=0} = 
\frac{1}{g^2}\, f_E\, \Lambda_s^3 \, ,
\label{Et}
\end{equation}
where $f_E\equiv f_E(\Lambda_s R) =0.21$--$0.26$.  Using this formula,
and assuming, in accordance with Ref.~\cite{Muell1}, the
$(N_c^2-1)/N_c$ dependence of the energy on the number of colors
$N_c$, we arrive at the values of 2700 GeV and of 25000 GeV for the
{\it initial} transverse energy per unit rapidity at RHIC and at LHC,
respectively~\cite{AlexRaj2}. We can also estimate a formation time
$\tau_f$ in our picture. For RHIC (LHC), we have $\tau_f = 0.3$ fm
($0.13$ fm). This then enables one to estimate the initial energy density 
to be $66.5$ GeV/fm$^3$ ($1.3$ TeV/fm$^3$) at RHIC (LHC).

The number and distribution of produced gluons are of considerable
interest as initial data for possible evolution of the gluon gas
towards thermal equilibrium~\cite{Muell1,BMSS,Jeff}.  Strictly
speaking, a particle number is only well-defined in a free-field
theory, and there is no unique extension of this notion to a general
interacting case.  For this reason, we use two different
generalizations of the particle number to an interacting theory, each
having the correct free-field limit. We verify that the two
definitions agree in the weak-coupling regime corresponding to late
proper times in the central region~\cite{AlexRaj3}.  Our first
definition is straightforward. We impose the Coulomb gauge condition
in the transverse plane: ${\vec\nabla}_\perp\cdot{\vec A}_\perp=0$ and
and determine the momentum components of the resulting field
configuration.  Our second definition is based on the behavior of a
free-field theory under relaxation. This ``cooling'' technique allows
a gauge invariant determination of the total particle number.
Unfortunately, it cannot be used to determine the momentum
distributions leaving the Colomb gauge determination as the only one
available.

We obtain
\begin{equation}
{1\over \pi R^2}\,{dN\over d\eta}|_{\eta=0} = {1\over g^2}\,
f_N \, \Lambda_s^2 \, ,
\label{BJN}
\end{equation}
where $f_N\equiv f_N(\Lambda_s L)= 0.13$ ($0.15$) for RHIC (LHC)
energies.  The agreement between the cooling and Coulomb techniques at
the larger values of $\Lambda_s L$ relevant for RHIC and LHC is
excellent.  It is not as good at the smaller values: in general, the
cooling number is more reliable.  Naively extrapolating our results to
SU(3), we find for $Au$-$Au$ central collisions at RHIC energies,
$dN/d\eta \sim 950$ for RHIC and $dN/d\eta \sim 4300$ for LHC
energies. 

Our results for the number distribution, computed in Coulomb gauge,
are as follows.  We have verified that for large $k_\perp$ our
numerically obtained multiplicity agrees with the lattice analogue of
the perturbative expression.  At smaller $k_\perp$, the distribution
softens and converges to a constant value, unlike its perturbative
counterpart.  Notably, this qualitative change of the distribution
occurs at $k_\perp\sim\Lambda_s$. We tried to quantify this
non-perturbative behavior by fitting the distribution to a variety of
physically motivated functional forms.  Surprisingly, we find that the
shape of the distribution is closely reproduced by the two dimensional
Bose-Einstein form $n(k)=A/(\exp(\beta\omega_k)-1)$, with the inverse
temperature $\beta$ of the order of 1 in units of $\Lambda_s$, and
with $\omega_k$ corresponding to a free {\it massive} dispersion relation,
with the mass of the order of $0.1\Lambda_s$.  This is a remarkable 
result for a purely classical theory, whose meaning, beyond providing
us with a convenient parametrization, is not yet clear.

We have studied the number distribution at different proper time
slices.  Our numerical results show clearly that the size of the mass
gap decreases as the square root of the proper time--this is what one
would expect for instance for a screening mass. Further, though the
peak of the distribution softens, its slope remains unchanged--the
inverse two dimensional ``temperature'' $\beta$ is a constant as a
function of proper time. Thus the particle number is essentially
unchanged--it increases slightly due to the ``decay'' of the mass
gap. 

Interestingly, all of these results (as well as the
$\Lambda_s/k_t$ form of the distribution for $k_t<\Lambda_s$) are
consistent with the theoretical analysis of Baier et al.~\cite{BMSS}
as communicated to us by Son~\cite{Son}. Baier et al. have studied the
importance of $2\rightarrow 3$ processes in driving the system to
equilibrium. The particular feature of their work, distinguishing it
from previous ones, is the inclusion of the
Landau--Pomeranchuk--Migdal (LPM) effect that results in the
suppression of hard gluon radiation in QCD~\cite{BDMPS}. In their kinetic
theory analysis, for proper times $\tau < 1/\alpha_S^{3/2}$, the
occupation number $f$ is greater than unity. This is the overlap
region between our classical simulations and kinetic theory 
and it is gratifying to see that they agree. This agreement further suggests 
that LPM interference effects are automatically included in the classical 
simulation.

What we have discussed above are the {\it initial} energy and number
distributions of gluons produced in nuclear collisions. The subsequent
evolution of the system (when the occupation number $f$ falls below
unity) is beyond the scope of the classical approach--it does however
provide the initial conditions for the evolution. The approach to equilibrium 
in the Mclerran--Venugopalan model was first investigated by 
Mueller~\cite{Muell1}. He initially considered only elastic scattering and 
derived a Landau--type equation for the single particle distributions. 
This equation has since been solved numerically and the approach to 
equilibrium studied quantitatively~\cite{Jeff}. One finds that elastic 
processes are very inefficient in driving the system to equilibrium (see 
also Ref.~\cite{Dumitru}). Baier et al. have argued that $2\rightarrow 3$ 
processes are not suppressed relative to $2\rightarrow 2$, and that they 
may actually be more efficient. For a more detailed discussion of 
thermalization, we refer the reader to their paper and to Son's talk at this 
conference~\cite{Son}.

\section{The CGC confronts RHIC data}

Data from RHIC were presented at this conference and some of it has 
already been published. How does the CGC picture fare when confronted 
with the data? The predictions of various models for the RHIC multiplicity 
have been summarized nicely by Eskola in his QM2001 talk~\cite{Eskola} 
(see also Ref.~\cite{Pajares}). Our prediction in Ref.~\cite{AlexRaj3} does 
quite well. This however was fortuitous since from Eq.~\ref{BJN} it is clear 
that the multiplicity depends very sensitively on the saturation scale--the 
latter clearly has at least a 10\% uncertainity.

A more sensitive test is the centrality dependence of the
multiplicity.  Kharzeev and Nardi plotted the charged particle
multiplicity per participant pair at pseudorapidity $\eta=0$ versus
the number of participants. They find that both the CGC and Glauber
models give nearly identical predictions~\cite{KharNar}. These in turn
agree well with the data~\cite{Roland}.

A consequence of the CGC picture is that $<p_t^2>\propto dN/dy/\pi R^2$, as is 
suggested by Eq.~\ref{BJN}. This dependence has been verified recently for 
the E735 $p\bar{p}$ data at the Tevatron as well as the UA(1) 
data~\cite{Juergen}. It also works for the NA49 Heavy Ion CERN SPS data. 
First comparisions with the RHIC data are also promising~\cite{privJuerg}.

In Ref.~\cite{AlexRaj3}, we found that the momentum distributions of the 
produced gluons were nearly universal functions that scaled as $f\equiv 
f(p_t/\Lambda_s)$. Whether the RHIC data reflects this can be easily checked 
because $\Lambda_s$ should be different for different centralities. 
Juergen Schaffner--Bielich has checked that the RHIC data for charged 
particles satisfies this property by rescaling the $p_t$ scale by 
$p_t\rightarrow \lambda p_t$, where $\lambda$ was an apparently 
arbitrary scaling constant. What is truly remarkable however is that the 
scaling constant is not arbitrary--it corresponds closely to the ratio of 
the saturation scales at the two centralities. If this scaling constant 
is used instead to determine $\Lambda_s$ at the different centralities and 
if then 
the number of charged particles per participant pairs at $\eta=0$ is plotted 
versus the number of participants (a la Kharzeev--Nardi), it agrees well with 
the data~\cite{privJuerg}.

So far so good. However, the ratio of $R_i=E_t/N_{\rm{gluon}}\sim 1.5
\Lambda_s$ computed by us~\cite{AlexRaj2,AlexRaj3} is too large
relative to the data~\cite{who}.  This is not necessarily a problem
because $R_i$ is not a conserved quantity and depends on the complex
dynamics of the space--time evolution. Our simulation only computes the 
{\it initial} value of this ratio. Both number changing processes
a la Baier et al. and the expansion of the system should decrease this ratio. 
If the former is the dominant process, one might be throwing the baby out 
with the bath water because the $\alpha_S$ dependence of the multiplicity 
might change. A more mundane possibility is that the function $f_E$ which 
was computed in SU(2) might be significantly lower for SU(3). That's one 
reason to press ahead with SU(3) simulations of the problem. As pointed 
out by Dumitru in his QM2001 talk, $R_i$ is a sensitive discriminant of 
the space--time dynamics in various models~\cite{Dumitru}. Detailed 
space--time simulations are in progress~\cite{Nara}.

Tests of the CGC scenario in the coming years include the energy 
dependence of $\Lambda_s$, event by event fluctuations, and heavy quark 
production among others. We have recently computed the likelihood of 
the formation of CP--odd domains in the classical approach~\cite{DAR}.

\section{Outlook}

The theoretical foundations of the classical approach to nuclear scattering 
clearly need to be on a firmer foundation. This is especially pressing in 
light of recent theoretical advances in small $x$ physics~\cite{ILM}. 
There are several practical 
improvements that need to be done independently of the above. 
The extension to SU(3) and relaxing the strict boost invariance assumption 
are on the top of the agenda. A related interesting problem is: does the 
hot gluonic matter thermalize? At least asymptotically, this problem should 
have a solution. Recent attempts to address this problem are encouraging 
but much work (hopefully aided by new ideas) remains. Finally, as more and 
more RHIC data pour in over the years, we expect that they will provide 
a sensitive test of the ideas discussed here. We live in interesting times!

\end{document}